# Prognostic Adjustment with Efficient Estimators to Unbiasedly Leverage Historical Data in Randomized Trials

**Lauren D. Liao**
ldliao@berkeley.edu
Division of Biostatistics, University of California, Berkeley

**Emilie Højbjerre-Frandsen**
ehfd@novonordisk.com
Biostatistics Methods and Outreach, Novo Nordisk A/S, and Department of Mathematical Sciences, Aalborg University

**Alan E. Hubbard**
hubbard@berkeley.edu
Division of Biostatistics, University of California, Berkeley

**Alejandro Schuler**
alejandro.schuler@berkeley.edu
Division of Biostatistics, University of California, Berkeley

**ABSTRACT**

Although randomized controlled trials (RCTs) are a cornerstone of comparative effectiveness, they typically have much smaller sample size than observational studies because of financial and ethical considerations. Therefore there is interest in using plentiful historical data (either observational data or prior trials) to reduce trial sizes. Previous estimators developed for this purpose rely on unrealistic assumptions, without which the added data can bias the treatment effect estimate. Recent work proposed an alternative method (prognostic covariate adjustment) that imposes no additional assumptions and increases efficiency in trial analyses. The idea is to use historical data to learn a prognostic model: a regression of the outcome onto the covariates. The predictions from this model, generated from the RCT subjects' baseline variables, are then used as a covariate in a linear regression analysis of the trial data. In this work, we extend prognostic adjustment to trial analyses with nonparametric efficient estimators, which are more powerful than linear regression. We provide theory that explains why prognostic adjustment improves small-sample point estimation and inference without any possibility of bias. Simulations corroborate the the-ory: efficient estimators using prognostic adjustment compared to without provides greater power (i.e., smaller standard errors) when the trial is small. Population shifts between historical and trial data attenuate benefits but do not introduce bias. We showcase our estimator using clinical trial data provided by Novo Nordisk A/S that evaluates insulin therapy for individuals with type II diabetes.

**Keywords**: *Causal inference, Diabetes, Historical Data, Prognostic Score, Randomized Trials*

# 1 Introduction

Practical, financial, and ethical concerns often preclude large randomized trials, which limits their power [1, 2, 3]. On the other hand, historical (often observational) data are often plentiful, and there are many existing methods for including historical data in trial analyses in order to boost power [4]. "Data fusion" methods simply pool trials with historical data [5, 6, 7]. Bayesian methods, which naturally rely on assumptions in the form of specified priors from historical data, are also popular in the literature [8, 9]. Similar problems have also been addressed in the generalizability and transportability research [6, 10, 11]. Recent studies proposed machine learning methods to integrate prior observational studies into trial analyses [12, 13].

Unfortunately, the aforementioned approaches are all sensitive to unobservable selection biases and must therefore be used with extreme care. The fundamental problem is that the historical population may differ systematically from the trial population in ways that impact both treatment assignment and outcome. For example, if the historical population did not have access to a modern standard-of-care, adding historical controls would artificially make any new drug seem more effective than it really is. Observable differences in populations can potentially be corrected under reasonable assumptions, but shifts in unobserved variables are impossible to detect or correct.

Schuler et al. (2021) take a different approach: instead of pooling data, they suggest using the historical data to train a *prognostic model* that predicts the outcome from baseline covariates [14] . They then adjust for the model's predictions on the trial data in the trial analysis using linear regression. They call this approach "prognostic adjustment".

There is no risk of bias because data is not pooled, yet efficicency can increase and be traded for smaller sample sizes. If there is a constant treatment effect and the outcome-covariate relationship is the same in the historical and trial populations, the proposed estimator even attains the semiparametric efficiency bound (more below). However, their method is limited to trial analyses using linear regression models.

Our task in this paper is to extend the prognostic adjustment approach beyond linear regression, specifically, to "semiparametrically efficient" estimators. Semiparametrically efficient estimators are those that attain the semiparametric efficient variance bound, which is the smallest asymptotic variance that any estimator can attain. The use of efficient estimators thus tends to reduce the uncertainty of the treatment effect estimate. These estimators leverage machine learning internally to estimate the treatment or the outcome model, or both; for example, the augmented inverse probability weighting estimator (AIPW) and the targeted maximum likelihood estimator (TMLE) are commonly used to evaluate the average treatment effect [15, 16, 17, 18, 19]. These estimators have been shown to improve the power of trials over unadjusted or linearly adjusted estimates [20].

In this study, we aim to improve power even further by incorporating historical data via prognostic adjustment. Our approach guarantees asymptotic efficiency of the trial treatment effect and more importantly, promises benefits in small-sample efficiency and robust inference.

## 2 Framework and notation

We follow the causal inference framework and roadmap from Petersen and van der Laan (2014) [21]. First, we define each observational unit $i \in \{1, ..., n\}$, as an independent, identically distributed random variable, $O_i$ with true distribution $P$. In our setting, each random variable $O = (W, A, Y, D)$ contains associated $p$ baseline covariates $W \in \mathbb{R}^p$, a binary treatment indicator $A$, denoting whether a unit is in the control group ($A = 0$), or in the treatment group ($A = 1$), an outcome $Y$, and an indicator $D$ denoting whether a unit is in either the trial ($D = 1$) or historical ($D = 0$) data sample.

We will assume that the trial data is generated under the setting of an RCT, such that $P(A = a|W, D = 1) = \pi_a$, with some positive constant $\pi_a$ denoting the treatment probability for $a \in \{0, 1\}$. Define $\mu_a(W) = E_P[Y|A = a, W, D = 1]$ as the conditional outcome means per treatment arm in the trial. Let $\rho_d(W) := E[Y \mid W, D = d]$ denote the *prognostic score* for a dataset $d$ [22]. When referenced without subscript ($\rho$) we are referring to the prognostic score in the historical data $D = 0$.

The fundamental problem of causal inference comes from not being able to observe the outcome under both treatment types. We know that for each individual, $Y = Y^A$, i.e., we observe the potential outcome corresponding to the observed treatment. To calculate the causal parameter of interest, we define the (unobservable) causal data to be $(Y^1, Y^0, A, W, D)$, generated from a causal data generating distribution $P^*$. In this study, we are interested in the causal average treatment effect (ATE) in the trial population:

$$\Psi^* = E_{P^*}[Y^1 - Y^0 \mid D = 1]. \tag{1}$$

which due to randomization in the trial is equal to the observable quantity:

$$\Psi = E_P[\mu_1(W) - \mu_0(W)|D = 1], \tag{2}$$

where $\mu_a(W) = E_P[Y|A = a, W, D = 1]$ is the conditional mean outcome in treatment arm $a \in \{0, 1\}$ from the observable data distribution.

Let $(\boldsymbol{W}, \boldsymbol{Y})$ denote a dataset with observed outcome $\boldsymbol{Y} = [Y_1, ..., Y_n]$ and the observed covariates $\boldsymbol{W} = [W_1, ..., W_n]$, where $(Y, W) \in (\mathbb{R} \times \mathbb{R}^m)$. Furthermore, let $\mathcal{L} : (\boldsymbol{W}, \boldsymbol{Y}) \mapsto f$ denote a machine learning algorithm that maps $(\boldsymbol{W}, \boldsymbol{Y})$ to a learned function $f$ that estimates the conditional mean $E[Y|W]$. The algorithm $\mathcal{L}$ may include detailed internal model selection and parameter tuning, and the algorithm works with predictors and data of any dimension (i.e., $m, n$ are arbitrary). Let $\widetilde{\boldsymbol{Y}}, \widetilde{\boldsymbol{W}}$ represent the historical dataset of size $\tilde{n}$, which is a draw from $P^{\tilde{n}}(Y, W \mid D = 0)$. We use $\hat{\rho}_0 = \mathcal{L}(\widetilde{\boldsymbol{Y}}, \widetilde{\boldsymbol{W}})$ (or just $\hat{\rho}$) to refer to an estimate of prognostic score learned from the historical data. Let $(\boldsymbol{Y}, \boldsymbol{A}, \boldsymbol{W})$ represent the trial data set of size $n$, which is a draw from $P^n(Y, A, W \mid D = 1)$. In a slight abuse of notation, let $\hat{\psi} = \hat{\psi}(\boldsymbol{Y}, \boldsymbol{A}, \boldsymbol{W})$ denote the mapping between trial data and our estimate $\hat{\psi}$ using an efficient estimator. For example, $\hat{\psi}$ could denote the cross-fit AIPW estimator described in Schuler (2021) [23].

## 3 Efficient estimators with prognostic score adjustment

Our proposed method for incorporating historical data with efficient estimators is simple: we first obtain a prognostic model by performing an outcome prediction to fit the historical data ($D = 0$) using a machine learning algorithm $\hat{\rho} = \mathcal{L}\left(\widetilde{\boldsymbol{Y}}, \widetilde{\boldsymbol{W}}\right)$. We then calculate the value of the prognostic score in the trial by feeding in all units' baseline covariates: $\boldsymbol{R} = \hat{\rho}(\boldsymbol{W})$. This prognostic score can be interpreted as a "pre-learned" dimension reduction for the trial covariates. Lastly, we estimate the ATE from the trial data, augmented with the prognostic score as an additional covariate, using an efficient estimator: $\hat{\psi}(\boldsymbol{Y}, \boldsymbol{A}, [\boldsymbol{W}, \boldsymbol{R}])$.

In practice, we suggest using a cross-validated ensemble algorithm (also called "super-learner") for $\mathcal{L}$ [24]. The super learner is known to perform as well as the best machine learning algorithm included in the library [24, 25]. The library in the super learner should include a variety of nonparametric and parametric learners, such as gradient boosting, random forest, elastic net, and linear models [25, 26].

For an efficient estimator, adding a fixed function of the covariates as an additional covariate will not change the asymptotic behavior [27, 28]. Thus our approach will never be *worse* than ignoring the historical data (as it might be if we pooled the data to learn the outcome regression). However, it also means that our approach cannot reduce asymptotic variance (indeed it is impossible to do so without making assumptions).

Nonetheless, we find that the *finite-sample* variance of efficient estimators is far enough from the efficiency bound that using the prognostic score as a covariate generally decreases the variance (without introducing bias) and improves estimation of the standard error. Mechanistically, this happens because the prognostic score "jump-starts" the learning curve of the outcome regression models such that more accurate predictions can be made with fewer trial data. This is especially true when the outcome-covariate relationship is complex and difficult to learn from a small trial. It is well-known that the performance of efficient estimators in RCTs is dependant on the predictive power of the outcome regression. Therefore improving this regression (by leveraging historical data) can reduce variance.

We expect finite-sample benefits as long the trial and historical populations and treatments are similar enough. But even if they are not identical, the prognostic score is still likely to contain very useful information about the conditional outcome mean.

## 3.1 Statistical properties

In this section, we theoretically show how adjusting for a prognostic score with an efficient estimator can improve estimation in a randomized trial. In an asymptotic analysis where the historical data grows much faster than the trial data, we show that using the prognostic score speeds the decay of the empirical process term in the stochastic decomposition of our estimator. The implications are that small-sample point estimation and inference should be improved even though efficiency gains will diminish asymptotically.

The material in this section assumes expertise with semiparametric efficiency theory and targeted/double machine learning. We present our results at a high level and do not present enough background for a casual reader. Two good starting points for this material are Schuler and van der Laan (2022) Ch.4 and Kennedy (2022) [29, 30]. The casual reader may skip this section and just review the "implications" below if they are comfortable with the above heuristic explanation of why prognostic adjustment may improve performance.

### 3.1.1 No asymptotic efficiency gain

Before showing how adjusting for a prognostic score for an efficient estimator can benefit estimation, we show that adding any prognostic score to an efficient estimator cannot improve asymptotic efficiency. To see this, we will start by considering the counterfactual means

$$\psi_a = E[E[Y|A=a, W]] = E[\mu_a(W)], \tag{3}$$

for any choice of $a \in \{0, 1\}$. We will return to the ATE shortly, but for now, it will make our argument clearer to only consider counterfactual means. Consider any efficient estimator for $\psi_a$ in a semiparametric model (known treatment mechanism) over the trial data $(Y, A, W)$. By definition, any efficient estimator of $E[\mu_a(W)]$ must have an influence function equal to the canonical gradient, which is referred to as the *efficient* influence function:

$$\varphi_a(Y, A, W) = \frac{1(A = 0)}{\pi_a}(Y - \mu_a(W)) + (\mu_a(W) - \psi_a), \qquad (4)$$

where $\pi_a = P(A = a)$ is the fixed, known propensity score. Consider now a distribution over $(Y, A, [W, R])$, where $R = g(W)$ for any fixed function $g$ playing the role of a prognostic model. The efficient influence function in this setting is the same as the above except $\mu_a(W) = E[Y|A = 0, W]$ is replaced by $\mu_a(W, R) = E[Y|A = 0, W, R]$. But since the prognostic score $R = g(W)$ is a fixed function of the covariates $W$, conditioning on the prognostic score after conditioning on $W$ does nothing, and we obtain $\mu_a(W, R) = \mu_a(W)$. Therefore, the prognostic score does not change the influence function and therefore cannot improve asymptotic efficiency. This holds even if you consider a random prognostic score learned from external data. The same holds for the ATE parameter since the efficient influence function in this case is $\varphi_{ATE} = \varphi_1 - \varphi_0$.

The fundamental issue is that the asymptotic efficiency bound cannot be improved without considering a different statistical model, e.g. distributions over $(Y, A, W, D)$. The problem is that in considering a different model we must also introduce additional assumptions to maintain identifiability of our trial-population causal parameter. For example, Li et al. (2021) consider precisely this setup and rely on an assumption of "conditional mean equivalence" $E[Y|A, W, D = 1] = E[Y|A, W, D = 0]$ to maintain identification while improving efficiency [13]. Similarly, an analysis following Chakrabortty, Dai, and Tchetgen Tchetgen (2022) shows that efficiency gains are also possible if we assume the covariate distributions are the same when conditioning on $D$ [31]. In this paper, we are interested in the case where we do not make these assumptions and therefore we have to look for benefits besides improvements in asymptotic efficiency. This is the content of the following sections.

### 3.1.2 Improving point estimation

We start by considering terms besides the efficient influence function term in the stochastic decomposition of our efficient estimator as compared to the efficient estimator adjusting for the prognostic score. As above we will consider estimation of the treatment-specific mean $\psi_a$, but from now on we will omit the $a$ subscripts to reduce visual clutter.

Consider the following decomposition:

$$\hat{\psi} - \psi = P_n \varphi + (P_n - P)(\hat{\varphi} - \varphi) + S - P_n \hat{\varphi} \qquad (5)$$

This sort of decomposition is common in the analysis of efficient estimators [29, 30]. As above, $\varphi$ here denotes the efficient influence function of $\psi$. We use the empirical process notation $P_n \hat{\varphi} = \frac{1}{n} \sum_{i=1}^{n} \hat{\varphi}(Y_i, A_i, W_i)$ to denote a sample mean, $P\hat{\varphi} = E[\hat{\varphi}(Y, A, W)]$ to denote a population mean (not averaging over randomness in $\hat{\varphi}$), and $\hat{\varphi}$ a plug-in estimate of $\varphi$ with $\mu$ estimated by regression. We will analyze each term and see what difference adjusting for the prognostic score makes in this general decomposition. Note that for an efficient estimator, the last term $-P_n \hat{\varphi} = 0$ by construction. Eliminating this "plug-in bias" term is the purpose of bias correcting schemes such as TMLE or efficient estimating equations [29, 30].

Of the remaining terms, the first is the efficient influence function term, $P_n \varphi$. We have already shown that an efficient estimator leveraging the prognostic score has the same influence function as one without. Therefore, this term is the same in either case.

Next we examine the remainder $S := [\hat{\varphi} - \varphi] + P\hat{\varphi}$. In the general setting of estimating a counterfactual mean in the nonparametric model, it is known that the remainder term $S$ can be bounded by the product of estimation errors in the outcome and propensity regressions $\|\hat{\mu} - \mu\| \, \|\hat{\pi} - \pi\|$ [29, 30]. But since the propensity score is known in a randomized trial, then there is no estimation error for that component and this term is exactly 0. Having a prognostic model does not change this and so again this term is identical with or without prognostic adjustment.

That leaves us with only the "empirical process" term $(P_n - P)(\hat{\varphi} - \varphi)$. With cross-fitting this term can be shown to be $O_P(\|\hat{\varphi} - \varphi\| n^{-\frac{1}{2}})$ [29, 30]. In our setting, the estimated influence function depends only on the estimated outcome regression and we thus have $(P_n - P)(\hat{\varphi} - \varphi) = O_P\big(\|\hat{\mu} - \mu\|n^{-\frac{1}{2}}\big)$. This is $o_P\big(n^{-\frac{1}{2}}\big)$ in all cases where the regression is $L_2$ consistent, but the actual rate *can be faster* depending on how quickly $\|\hat{\mu} - \mu\|$ converges. We'll use $t_n$ to denote this rate, i.e. $\|\hat{\mu} - \mu\| = o_P(t_n)$.

For example, presume we know the dimensionality of our covariates is $p = 16$ that our unknown function $\mu$ has at least $b = 2$ derivatives. By Stone (1982), the optimal rate for any nonparametric regression estimator $\mathcal{L}$ is $t_n = n^{-\frac{1}{2+p/b}} = n^{-\frac{1}{10}}$ and by usual "$o_P$ algebra" (see van der Vaart (2000) Ch. 2) the empirical process would be $o_P\big(t_n n^{-\frac{1}{2}}\big) = o_P\big(n^{-\frac{1}{10}} n^{-\frac{1}{2}}\big) = o_P\big(n^{-\frac{6}{10}}\big)$ [32, 33]. If we could instead somehow attain a faster $t_n = n^{-\frac{1}{5}}$ rate on the convergence of the outcome regression, our empirical process rate would be the faster $n^{-\frac{7}{10}}$. This is the mechanism by which prognostic adjustment improves efficient estimators. When the prognostic score is used to estimate $\hat{\mu}$, the norm $\|\hat{\mu} - \mu\|$ is generally smaller than it otherwise would be.

We can formalize this asymptotically. Recall that $n$ denotes the trial sample size and $\tilde{n}$ denotes the historical sample size. Consider a historical sample much larger than the trial in the limit: $\frac{n}{\tilde{n}} = r_n \to 0$. For example, presume $\tilde{n} = n^2$ in which case $r_n = \frac{1}{n}$. Presume that the historical distribution of covariates and outcome is the same as that in the trial control arm, or simply that $\rho(w) = \mu_0(w)$ (this is a best-case scenario). Presume we have a learning algorithm $\mathcal{L}$ that in some large nonparametric function class can learn functions in an $L_2$ sense at rate $t_n$ (e.g. $n^{-\frac{1}{10}}$ as above). Let $\hat{\rho}_{\tilde{n}}$ be the prognostic score learned from the $\tilde{n}$ historical samples using this learner.

Instead of fitting our trial outcome regression with the prognostic score as a covariate, presume that we directly take $\hat{\mu}_{n,0} = \hat{\rho}_{\tilde{n}}$. In other words, we use our prognostic model as the control outcome regression in the trial, ignoring the trial data. That's sensible in this hypothetical setting because 1) the prognostic model will indeed converge to the true outcome function and 2) the amount of omitted trial data is vanishingly small relative to the historical data. We can also interpret this as an estimator of $E[Y|A = 0, W, \rho_{\tilde{n}}(W)]$ in the trial data (i.e. our prognostically adjusted outcome regression): we should expect that when the prognostic model is good, our learner will simply return the prognostic score untouched to model the control counterfactual outcome. Or, more formally, if a parametric learner is used on top of the prognostic score (e.g. a working model $Y = \beta_0 + \hat{\rho}(W)$) then the total error will be composed of a root-$n$ term from learning the parameters and a higher order empirical process term.

If we used our learner to fit the outcome regression from trial data alone, our rate on the $L_2$ norm of the outcome regression would be $t_n$, e.g. $n^{-\frac{1}{10}}$ in our example. However, if we use the prognostic score the rate is $t_{\tilde{n}}$ ($\tilde{n}$ instead of $n$), e.g. $\tilde{n}^{-\frac{1}{10}}$. But recalling $\frac{n}{\tilde{n}} = o(r_n)$ ($r_n = \frac{1}{n}$, for example), we can express $t_{\tilde{n}}$ as $\frac{t_n r_n}{n}$. Essentially we can plug in $\tilde{n} = n^2$ into $t_{\tilde{n}} = \tilde{n}^{-\frac{1}{10}}$ to see $t_n = (n^2)^{-\frac{1}{10}} = n^{-\frac{1}{5}}$. This can dramatically increase the speed at which $\|\hat{\mu} - \mu\| \to 0$ in terms of $n$ and consequently affect the rate of convergence of the empirical process term, making it decay faster than it would without the prognostic adjustment.

The result of making the empirical process term higher order is to reduce finite-sample variance of our point estimate (i.e. improve finite-sample efficiency). The reason for this is that with cross-fitting the empirical process term is already exactly mean-zero [29, 30]. Therefore, improving its convergence rate must only affect the variance of our estimate and not its bias.

### 3.1.3 Improving standard error estimation

We can apply similar arguments to show that performance of the plug-in estimate of asymptotic variance $P_n \hat{\varphi}^2$ (based on our estimated influence function $\hat{\varphi}$) is also improved by prognostic adjustment. The difference between the estimate and the true asymptotic variance $P\varphi^2$ can be decomposed as

$$P_n \hat{\varphi}^2 - P\varphi^2 = (P_n - P)\varphi^2 + (P_n - P)(\hat{\varphi}^2 - \varphi^2) + P(\hat{\varphi}^2 - \varphi^2) \qquad (6)$$

The first term here is a nice empirical mean which by the central limit goes to zero at a root-$n$ rate and which is unaffected by prognostic adjustment.

The second term is similar to the empirical process term discussed above in the context of point estimation and by identical arguments this term decays faster when prognostic adjustment is used (note $L_2$ convergence of $\hat{\varphi}$ implies the same for $\hat{\varphi}^2$ under regularity conditions that are satisfied in our setting because $\pi$ is a known constant bounded away from 0). This term is always higher-order than $n^{-\frac{1}{2}}$ and thus asymptotically negligible, but possibly impactful in small samples. It is also mean-zero. Together, this means that its improved rate with prognostic adjustment translates to less finite-sample variability in our estimate of the standard error.

The last term is bounded by $\|\hat{\varphi}^2 - \varphi^2\|$ so this term also decays faster with prognostic adjustment. This term contributes to both bias and variance of the standard error estimate. Unlike the equivalent norm that appears in the bound for the empirical process term, the norm here is not divided by $\sqrt{n}$ and so this term may be of *leading order* (slowest decaying) in the overall stochastic decomposition and thus asymptotically relevant. Since prognostic adjustment increases the rate, the term may go from leading order to higher-order. Therefore prognostic adjustment may, in some cases, make the plug-in standard error an asymptotically linear estimator (i.e. standard error of the standard error should decrease at a $\frac{1}{\sqrt{n}}$ rate).

### 3.1.4 Caveats

Although we do not need additional assumptions for identifiability and thus retain unbiased estimation in all cases, all of the possible benefits described above do rely on the assumption that the historical and trial data-generating processes share a control-specific conditional mean. If this is not the case, then the amount by which the prognostic score speeds convergence of the control outcome regression will be attenuated, but not necessarily eliminated. For example, if the true outcome regression is the same as the prognostic score up to some parametric transformation that is learnable at a fast rate by a learner in our library $\mathcal{L}$, then we should still expect benefits.

Until now we have also focused on the control counterfactual mean. The influence function for the ATE is the difference of those for the two counterfactual means and consequently we can decompose the empirical process term into two terms which are $O_P\big(\|\hat{\mu}_1 - \mu_1\| \, n^{-\frac{1}{2}}\big)$ and $O_P\big(\|\hat{\mu}_0 - \mu_0\| \, n^{-\frac{1}{2}}\big)$ where $\mu_a = E[Y|A=a, W]$ denote the counterfactual mean functions. Therefore, the overall order of the empirical process term is dominated by whichever of these terms is lower-order. For the treatment regression to leverage the historical data, we need to assume some prognostic information can be transferred from the control to the treatment arm. For example, we might expect satisfactory information transfer from the historical data to the trial treatment arm when when there is a constant treatment effect $c$ (i.e. $\mu_1(w) = \mu_0(w) + c$) or there is some other parametric, easily-learnable relationship between the two conditional means. Otherwise, the slower convergence rate for $\mu_1$ dominates and we obtain less benefit from prognostic adjustment. A worst-case scenario would be when the two conditional means depend on mutually exclusive sets of covariates: if this is the case, no transfer should be possible and benefits should be limited or absent.

### 3.1.5 Implications

Our analysis shows that use of the historical sample via prognostic score adjustment produces less-variable point estimates in small samples as well as more stable and accurate estimates of standard error. Unfortunately, asymptotic gains in efficiency are not possible without further assumptions.

However, these benefits are contingent on the extent to which the covariate-outcome relationships in both treatment arms of the trial are similar to the equivalent relationship in the histoircal data. In particular, differences between historical and trial populations and high heterogeneity of effect may both attenuate benefits. This is also the case with existing Bayesian or pooling approaches that incorporate historical data into trial analyses. Nonetheless, these problems can never induce bias. Therefore, relative to alternatives, prognostic adjustment of efficient estimators provides strict guarantees for type I error, but at the cost of limiting the possible benefits of using historical data.

## 4 Simulation study

This simulation study aims to demonstrate the utility of an efficient estimator with the addition of a prognostic score. We examine how our method performs in different data generating scenarios (e.g. heterogeneous vs. constant effect), across different data set sizes, and when there are distributional shifts from the historical to the trial population. The simulation code is publicly online at https://github.com/ldliao/tl4rct.

## 4.1 Setup

The simulation study is based on the structural causal model in Equation 7. In total there are 20 observed covariates of various types and a single *unobserved* covariate.

$$\begin{aligned}
W_1 &\sim \text{Unif}(-2, 1) \\
W_2 &\sim \mathcal{N}(0, 3) \\
W_3 &\sim \text{Exp}(0.8) \\
W_4 &\sim \Gamma(2, 1) \\
W_6 ... W_{20} &\sim \text{Unif}(0, 1) \\
U &\sim \text{Unif}(0, 1) \\
A &\sim \text{Bern}\left(\frac{1}{2}\right) \\
Y^a | W, U &= m_a(W, U) + \mathcal{N}(0, 2) \\
Y &= AY^1 + (1 - A)Y^0
\end{aligned} \quad (7)$$

Notice that $m_a(W, U)$ is the mean of the counterfactual conditioned on both the observed and unobserved covariates. Our observable conditional means are thus $\mu_a(W) = E[m_a(W, U) | W]$. We examine two different scenarios for the conditional outcome mean $m_a$. In our "heterogeneous effect" simulation:

$$\begin{aligned}
m_1(W, U) &= (10 \cdot \sin(|W_1|\pi))^2 + 1(U > 1.01) \cdot 8 + 1(U > 1.55) \cdot 15 - 42 \\
m_0(W, U) &= 10 \cdot \sin(|W_1|\pi) + 1(U > 1.01) \cdot 8 + 1(U > 1.55) \cdot 15
\end{aligned} \quad (8)$$

and in our "constant effect" simulation:

$$\begin{aligned}
m_1(W, U) &= 10 \cdot \sin(|W_1|\pi) + 1(U > 1.21) \cdot 20 + 1(U > 1.55) \cdot 15 - 0.8 \\
m_0(W, U) &= 10 \cdot \sin(|W_1|\pi) + 1(U > 1.21) \cdot 20 + 1(U > 1.55) \cdot 15
\end{aligned} \quad (9)$$

To begin, we use the same DGP for the historical and trial populations except the fact that $A = 0$ determinstically in the historical DGP. But in what follows, we loosen this assumption by changing the historical data generating distribution with varying degrees of observed and unobserved covariate shifts.

We examine several scenarios: first, we analyze the trial ($n = 250$) under the heterogeneous and constant treatment effect DGPs, where the historical sample ($\tilde{n} = 1,000$) is from the same DGP (Equation 7) as the trial sample. Second, we vary the historical and trial sample sizes for the heterogeneous treatment effect simulation. To vary the historical sample sizes, we first fix the trial sample size ($n = 250$) and vary the historical sample size (with $\tilde{n} = 100, 250, 500, 750$, and $1,000$). To vary the trial sample sizes, we first fix the historical sample size ($\tilde{n} = 1,000$) and vary the trial sample size (with $n = 100, 250, 500, 750$, and $1,000$). We also vary $n$ and $\tilde{n} = n^2$ together to demonstrate asymptotic benefits in the estimation of the standard error (as discussed in section 3).

Third, we examine the effect of distributional shifts between the historical and trial populations. In these cases we draw trial data from the DGP in Equation 7 but draw our historical data from modified versions. To simulate a "small" observable population shift we let $W_1|D = 0 \sim \text{Unif}(-5, -2)$ and to simulate a "large" observable population shift we let $W_1|D = 0 \sim \text{Unif}(-7, -4)$. To simulate a "small" *unobservable* population shift we let $U|D = 0 \sim \text{Unif}(0.5, 1.5)$ and to simulate a "large" *unobservable* population shift we let $U|D = 0 \sim \text{Unif}(1, 2)$. The shifts in the unobserved covariate induce shifts in the conditional mean relationship between the observed covariates and the outcome (see Supplemental Appendix B for an explicit explanation).

We consider three estimators for the trial: unadjusted (difference-in-group-means), linear regression (with $\text{HC}_3$ robust standard error estimator [34, 35, 36]), and targeted maximum likelihood estimation (TMLE; an efficient estimator [15]). All estimators return an effect estimate and an estimated standard error, which we use to construct Wald 95% confidence intervals and corresponding p-values. The naive unadjusted estimator cannot leverage any covariates, but both linear and TMLE estimators can. We compare and contrast results from linear and TMLE estimators both with and without the fitted prognostic score as an adjustment covariate ("fitted") to compare against Schuler et al. [14]. We also consider the oracle version of the prognostic score ("oracle") for a benchmark comparison; the oracle prognostic score perfectly models the expected control outcome in the trial $E[Y|W, A = 0, D = 0]$. Unlike the fit prognostic score, the oracle version is not affected by random noise in the historical data and it is not sensitive to shifts between historical and trial populations (indeed it is not affected by the historical data at all). The oracle prognostic score only serves as a best-case comparison and is infeasible to calculate in practice.

For simplification, we include the same specifications of the discrete super learner (cross-validated ensemble algorithm) for both the prognostic model and all regressions required by our efficient estimators. Specifically, we use the discrete super learner – choosing one machine learning algorithm from a set of algorithms for each cross-fit fold via the lowest cross-validated mean squared error. The set of algorithms include the linear regression, gradient boosting with varying tree tuning specifications (xgboost) [37], and Multivariate Adaptive Regression Splines [38]. Specifications for tuning parameters are in Supplemental Appendix A.

We repeat all simulation scenarios 200 times to calculate all performance metrics. We calculate the average estimated standard error, empirical power (percentage of significant p-values, i.e., Wald CIs not covering the true ATE), estimation root mean square error (RMSE), and coverage.

## 4.2 Results

Our results for the heterogeneous effect scenario are summarized in Table 1 and Table 2. Results for other DGPs are qualitatively similar so these are reported in Supplemental Appendix C along with additional performance metrics.

| Prognostic Score | TMLE | linear | unadjusted |
|---|---|---|---|
| oracle | -0.069 (4.341) | 0.034 (9.727) | - |
| fit | -0.064 (4.482) | 0.025 (9.710) | - |
| none | 0.009 (5.691) | 0.072 (9.774) | 0.153 (9.509) |

Table 1: Empirical bias and variance for the point estimate of the trial ATE in the heterogeneous effect simulation scenario. Results in the table are formatted as "empirical bias (empirical variance)"

Table 1 illustrates the mean of the empirical bias and empirical variance of the ATE point estimate across the 200 simulations. The results demonstrate that prognostic adjustment decreases variance relative to vanilla TMLE in the realistic heterogeneous treatment effect scenario (results are similar in other scenarios). The reduction in variance results in an increase in an 11% increase in power in this case. In terms of variance reduction, fitting the prognostic score is almost as good as having the oracle in this most scenarios.

Prognostic adjustment also improves the variance of the linear estimator (corroborating Schuler (2021)) [14]. But overall TMLE convincingly beats the linear estimator, with or without prognostic adjustment, except for in the constant effect scenario where the two are roughly equivalent with prognostic adjustment. The matching or slightly superior performance of prognostically adjusted linear regression in the constant effect DGP is consistent with the optimality property previously discussed in Schuler et al. (2021) [14].

Importantly, the variance is not underestimated in any of our simulations meaning that the coverage was nominal (95%) for all estimators (and thus strict type I error control was attained; Supplemental Appendix C). Including the prognostic score did not affect coverage in any case, even when the trial and historical populations were different.

| Prognostic Score | TMLE | linear | unadjusted |
|---|---|---|---|
| oracle | 0.152 (0.003) | 0.138 (0.028) | - |
| fit | 0.158 (0.005) | 0.137 (0.029) | - |
| none | 0.180 (0.055) | 0.213 (0.026) | 0.139 (0.011) |

Table 2: Empirical bias and variance for the estimated standard error of the trial ATE estimate. Results in the table are formatted as "empirical bias (empirical variance)"

Table 2 illustrates the mean of the empirical bias and empirical variance of the *estimated standard error* of the ATE estimate. The table corroborates the theoretical findings from Section 3, namely that the variance of the estimated variance for an efficient estimator (TMLE) is decreased by prognostic adjustment.

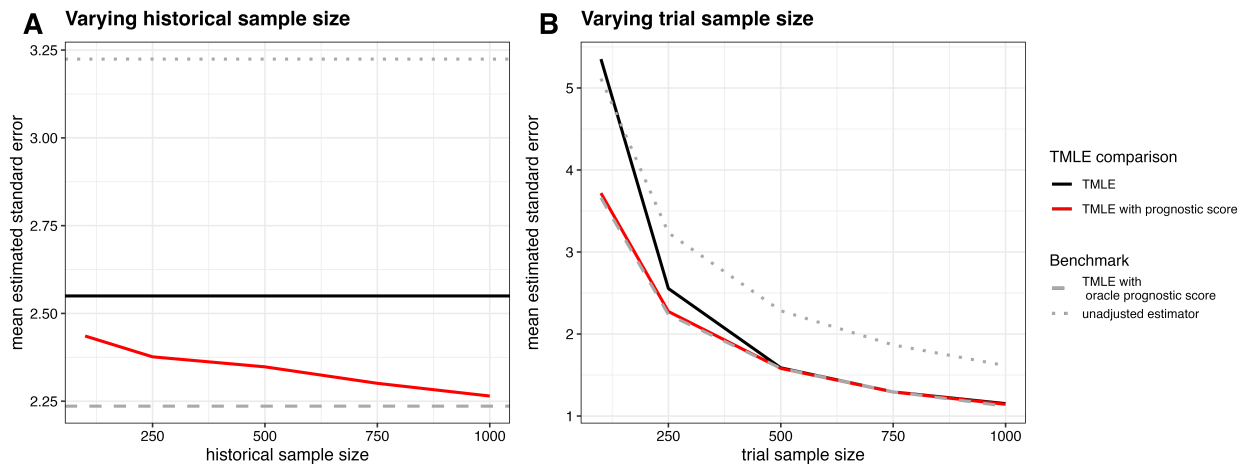

Figure 1: Mean estimated standard errors across estimators when historical and trial sample sizes are varied using the heterogeneous DGP. When the historical sample size is varied (Figure 1.A), the trial is fixed at $n = 250$. When the trial size is varied (Figure 1.B), the historical sample is fixed at $\tilde{n} = 1000$.

Using larger historical data sets increases the benefits of prognostic adjustment with efficient estimators. Figure 1.A shows a detailed view of this phenomenon in terms of decrease in the average estimated standard error as the historical data set grows in size. In effect, the larger the historical data, the smaller the resulting confidence intervals tend to be in the trial (while still preserving coverage, see Supplemental Appendix C), for the estimators leveraging an estimated prognostic score. Figure 1.B shows the change in estimated standard error as the *trial* size varies. This illustrates that the relative benefit of prognostic adjustment is larger in smaller trials. Here we see an 11% increase in power comparing the TMLE with versus without fitted prognostic score when $n = 250$, but an 80% increase when $n = 100$. From Figure 1 we again see that the TMLE with the fitted prognostic score performs almost as well as the TMLE with the oracle prognostic score when the historical sample size is increased to around 1000.

When trial sample size $n$ and historical sample size $\tilde{n} = n^2$ increase together, our theory predicts that the plug-in standard error may become asymptotically linear with prognostic adjustment. This is confirmed by simulation (with 1,000 Monte Carlo simulations): as we increase $n$ with prognostic adjustment, the empirical variance of the estimated standard error times $n$ is closer to a flat line which indicates a near-$\sqrt{n}$ rate of decay.

The same is not true without prognostic adjustment: the variance of the plug-in standard error from TMLE is greater and falls more slowly.

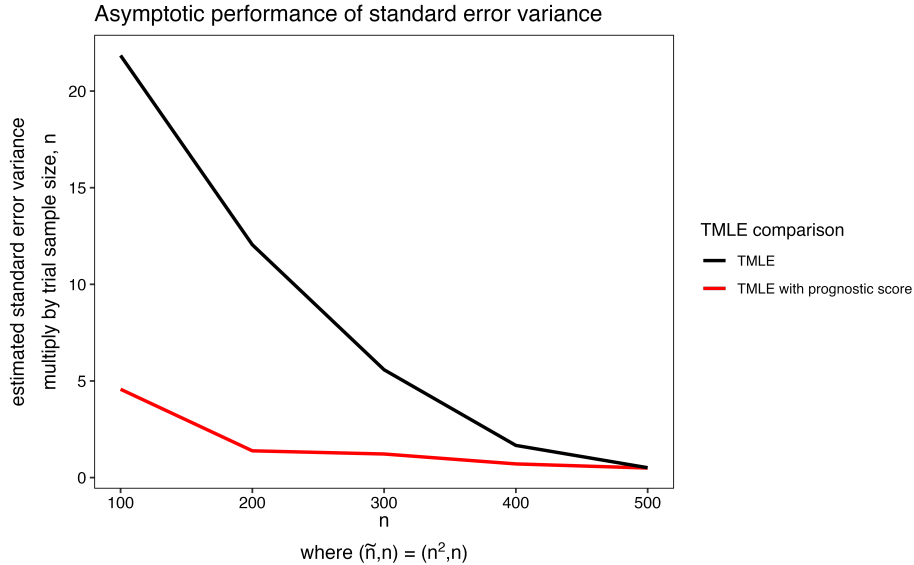

Figure 2: Variance of estimated standard error across estimators when historical and trial sample sizes are varied using the constant effect DGP. The historical sample size, $\tilde{n}$ is proportional to trial size $n$, where $(\tilde{n}, n) = (n^2, n)$.

We also observe that our method is relatively robust to both observed and unobserved distributional shifts between historical and trial populations (Figure 3). When the shifts are large, the prognostic score may be uninformative (most evident in subplot (3.B)), but including it may still improve efficiency (as seen in subplot (3.A)). We also see that a good prognostic score (no shift in distribution) substantially reduces the variability of the estimated standard error. Variability increases with the magnitude of the covariate shift but still does not exceed that of TMLE without prognostic adjustment.

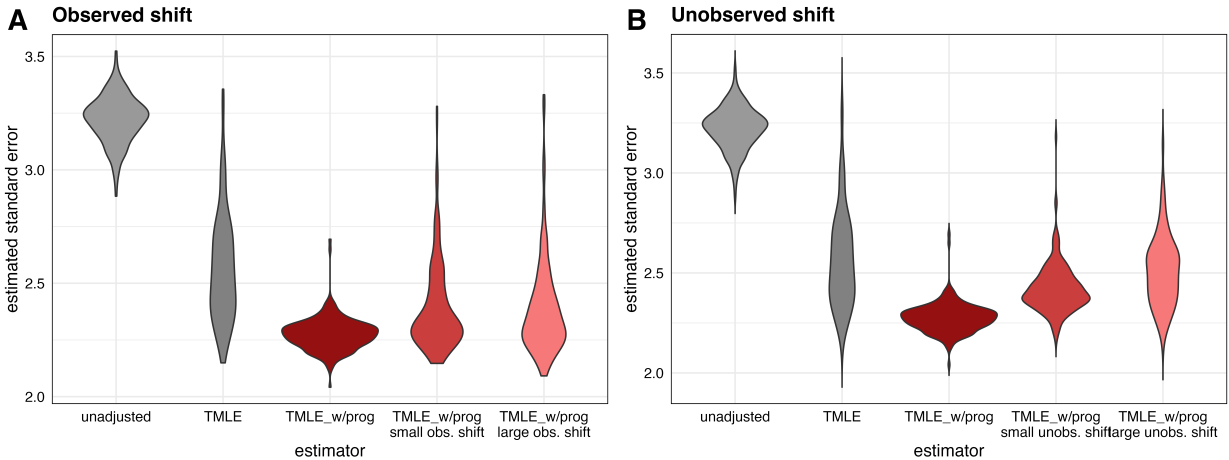

Figure 3: Estimated standard errors across estimators when observed (Figure 3 .A) and unobserved shifts (Figure 3 .B) are present in the historical sample relative to the trial sample.

## 5 Case Study

In this section, we examine the use of TMLE with prognostic covariate adjustment in RCTs involving people diagnosed with type 2 diabetes (T2D). T2D is a chronic disease with a progressive deterioration of glucose control. Glucose control is normally evaluated by long-term blood glucose level, measured by hemoglobin A1C (HbA1C). The analyses are carried out using data provided by Novo Nordisk A/S originating from 14 previously conducted RCTs within the field of diabetes, see Supplemental Appendix D for a full overview of the trials.

We reanalyse the phase IIIb clinical trial called NN9068-4229, where the trial population consisted of insulin naive people with T2D [39]. The participants of this trial were inadequately controlled on treatment with SGLT2i, a type of oral anti-diabetic treatment (OAD). Inadequately controlled was defined as having a HbA1C of $7.0 - 11.0\%$ (both inclusive). The aim of the trial was to compare glycaemic control of insulin IDegLira versus insulin IGlar as add-on therapy to SGLT2i in people with T2D. The trial was a 26-week, 1:1 randomized, active-controlled, open label, treat-to-target trial with 420 enrolled participants. One participant was excluded due to non-exposure to trial product, yielding $n = 419$. The efficacy of IDegLira was measured by the difference in change from baseline

HbA1C to landmark visit week 26. Our corresponding historical sample came from previously conducted RCTs with a study population also consisting of insulin naive people with T2D, who were inadequatly controlled on their current OADs. A total of $\tilde{n} = 3311$ participants all receiving insulin IGlar, were enrolled in the historical sample.

For the trial reanalysis in our study, we included patient measures of their demographic background, laboratory measures, concomitant medication, and vital signs. The treatment indicator where only used in the NN9068-4229 trial. For details on the specific measurements, covariate distributions, and imputation of missing covariates see Appendix E, F, and G. For the continous covariates we see that the mean and standard deviation are not particularly different between the historical and new trial sample, meaning that both resemble a T2D population with uncontrolled glycaemic control. Furthermore we see that the range of continous covariates for the new trial sample are contained in the range of the historical sample. This indicates that the trial population is largely similar to the historical population, at least in terms of observable covariates. For the categorical covariates the distributions vary between the historical and new trial sample. However, all the categories in the new trial sample are present in the historical sample.

An ANCOVA model (linear estimator) with baseline HbA1C, region and pre-trial OADs as adjustment covariates was used in the original analysis of the primary endpoint in the NN9068-4229 trial. In this analysis, an average treatment effect estimate of -0.340 (95% confidence interval [-0.480;-0.200]). In this reanalysis, we report the result of five estimators: unadjusted, linear regression (adjusting for all available covariates), linear regression with a prognostic score, TMLE, and TMLE with a prognostic score. For this application, we expanded the library of the super learner for a more comprehensive set of machine learning models than the simulation, including random forest [40], k-nearest neighbor, and a more comprehensive set of tuning parameters for the xgboost model in addition to the previously specified library, see Supplemental Appendix A. Separately, we obtained the correlation of the fitted prognostic score against the trial outcome. The prognostic score's correlation with the outcome is 0.752 with control subjects and 0.622 with treated subjects, indicating that adjustment for the score should result in an improvement over unadjusted estimation [14].

The results of the reanalysis are shown in Table 3.

|  | **TMLE** | **Linear** | **Unadjusted** |
|---|---|---|---|
| With prognostic score | $-\mathbf{0.351}$ (**s.e. 0.145**) | $-0.355$ (s.e. 0.157) |  |
| Without prognostic score | $-0.369$ (s.e. 0.150) | $-0.355$ (s.e. 0.157) | $-0.248$ (s.e. 0.192) |

Table 3: Estimates for average treatment effect (ATE) and with 95% confidence levels (CL) of change in hemoglobin A1C (HbA1C) from baseline to week 26 for insulin IDegLira versus insulin IGlar as add-on therapy to SGLT2i in people with T2D. This is a reanalyses of the NN9068-4229 trial using five different estimators.

From Table 3, we see that the smallest confidence interval is obtained using TMLE with prognostic score. All methods obtain similar point estimates except from the unadjusted estimator. Notice that the linear estimator with or without a prognostic score yields the same results, since the prognostic model is a linear model in this case (chosen as the model that yielded lowest MSE from 20-fold cross-validation within the discrete super learner).

As illustrated by the simulation study and the assymptotic analysis in Section 3.1, the relative benefit of prognostic adjustment is larger in smaller trials. To examine this result, we subsampled from the the NN9068-4229 trial but reanalyzed with selecting 50 participants randomly from each group, resulting in $n = 100$. This random selection of 50 participants from each group is repeated 10 times and averaged to compute the point estimate and standard error. The average correlation of the prognostic score with the outcome was 0.790 with control subjects and 0.656 with treated subjects. The results of the reanalysis is shown in Table 4.

|  | **TMLE** | **Linear** | **Unadjusted** |
|---|---|---|---|
| With prognostic score | **−0.519 (s.e. 0.307)** | −0.544 (s.e. 0.438) | |
| Without prognostic score | −0.582 (s.e. 0.349) | −0.544 (s.e. 0.438) | −0.344 (s.e. 0.399) |

Table 4: Estimates for average treatment effect (ATE) and with 95% confidence levels (CL) of change in hemoglobin A1C (HbA1C) from baseline to week 26 for insulin IDegLira versus insulin IGlar as add-on therapy to SGLT2i in people with T2D. This is a reanalyses of the NN9068-4229 trial using five different estimators where 50 participants from the control and treatment group, respectively, were chosen at random yielding a total sample size of $n = 100$. The random selection is done 10 times and the reported numbers are the average of the point estimates and standard error.

We see an relatively larger reduction in the standard error estimate using TMLE with prognostic covariate adjustment compared to TMLE without in the reanalysis.

# 6 Discussion

In this study we demonstrate the utility of incorporating historical data via a prognostic score in an efficient estimator while maintaining strict type I error control. Using the prognostic score via covariate adjustment overall improves the performance of the efficient estimator by decreasing the standard error and improving its estimation. This method is most useful in randomized trials with small sample sizes. Our proposed method is shown to be robust against bias even when the historical sample is drawn from a different population.

Prognostic adjustment requires no assumptions to continue to guarantee unbiased causal effect estimates. However, this comes with a tradeoff: without introducing the risk of bias, there is a limit on how much power can be gained and in what scenarios. For example, the method of Li et al. (2021) (which imposes an additional but rather light assumption) can asymptotically benefit from the addition of historical data, whereas our method can only provide gains in small samples [13]. However, these gains are *most important* precisely in small samples because estimated effects are likely to be of borderline significance, whereas effects are more likely to be clear in very large samples regardless of the estimator used.

Besides being assumption-free, our method has other practical advantages relative to data fusion approaches. For one, we do not require a single, well-defined treatment in the historical data. Moreover, we do not require an exact overlap of the covariates measured in the historical and trial data sets, as long as there is some overlap and informativeness, the historical data can be leveraged.

It is also easy to utilize multiple historical data sets: if they are believed to be drawn from substantially different populations, separate prognostic scores can be built from each of them and included as covariates in the trial analysis. As long as one of these scores is a good approximation of the outcome-covariate relationship in one or more arms of the trial, there will be added benefits to power. The addition of multiple covariates poses no risk for efficient estimators that use data-adaptive machine learning methods, which is another advantage of marrying prognostic adjustment to efficient estimation as opposed to marrying it to traditional parametric adjustment.

Prognostic adjustment with efficient estimators can also be used with pre-built or public prognostic models: the analyst does not need direct access to the historical data if they can query a model for predictions. This is helpful in cases where data is "federated" and cannot move (e.g. when privacy must be protected or data has commercial value). Specifically, individual subject data is not necessary to perform this step.

The theory we developed to explain the benefits of prognostic adjustment in the context of efficient estimation for trials is easily generalizable to estimation of any kind of pathwise differentiable parameter augmented with transfer learning from an auxiliary dataset. The specific breakdown of different terms may differ but the overall intuition should be the same: transfer learning may accelerate the disappearance of higher-order terms that depend on the error rates of regression estimates.

Our approach is closely related to the transfer learning literature in machine learning. In transfer learning, the goal is to use a (large) "source" dataset to improve prediction for a "target" population for which we have only minimal training data [41, 42, 43]. In this work we use a particular method of "transfer" (adjusting for the source/historical model prediction) to improve the target (trial) predictions, which drives variance reduction. It should also be possible to leverage other more direct forms of model transfer for the out-come regression, such as pre-training a deep learning model on the historical data and then fine-tuning using the trial data.

Lastly, since we use efficient estimators, we can leverage the results of Schuler (2021) to prospectively calculate power with prognostic adjustment [23]. In fact, we suspect the methods of power calculation described in that work would improve in accuracy with prognostic adjustment since the outcome regressions are "jump-started" with the prognostic score. Verification of this fact and empirical demonstration will be left to future work.

# Acknowledgments

The authors would like to thank study participants and staff for their contributions. This research was conducted on the Savio computational cluster resource provided by the Berkeley Research Computing program at the University of California, Berkeley. This computing resource was supported by the UC Berkeley Chancellor, Vice Chancellor for Research, and Chief Information Officer. The authors thank Christopher Paciorek for answering Savio related inquiries.

# Data availability statement

Data are not available for this study.

# Funding

This research was made possible by funding from the National Science Foundation (DGE 2146752) to LDL and global development grant (OPP1165144) from the Bill & Melinda Gates Foundation for AEH to the University of California, Berkeley, CA, USA. This research also received funding from Innovation Fund Denmark (Grant number 2052-00044B) for EHFD to Novo Nordisk A/S. The funders had no role in study design, data collection and analysis, decision to publish, or preparation of the manuscript.

# Financial disclosure

None reported.

# Conflict of Interest

The authors declare no potential conflict of interests.

# Appendix A: Discrete learner specifications for simulation and case study

**Discrete super learner specifications**

Machine learning is performed through discrete super learner that the targeted maximum likelihood estimator internally leverages. For simplicity, the prognostic model is built using the discrete super learner as well. A discrete super learner selects from a set of candidate models (i.e., the library) to obtain a single, best prediction model via cross-validation. In this section, we describe the exact tuning parameters and set up for the simulation and case study.

**Simulation set up**

    **Cross-validation** 5-fold cross-validation is used to select the best candidate learner in the library for historical sample size 1,000, and 10-fold cross-validation for historical sample size less than 1,000.

    **Cross-fit** 5-fold Cross-fitting is employed.

    **Discrete super learner library**

    Multivariate Adaptive Regression Splines with the highest interaction to be to the 3rd degree, linear regression, extreme gradient boosting with specifications: learning rate 0.1, tree depth 3, crossed with trees specified 25 to 500 by 25 increments. Cases with *fitted* prognostic score include an augmented library that includes candidate learners with prognostic score in addition.

    **Discrete super learner specifications** loss function is specified to be the mean square error loss.

**Case study set up**

    **Cross-validation** 20-fold cross-validation is used to select the best candidate learner in the library.

    **Cross-fit** 20-fold Cross-fitting is employed.

    **Discrete super learner library**

    Multivariate Adaptive Regression Splines with the highest interaction to be to the

3rd degree, logistic regression, extreme gradient boosting with specifications: learning rate 0.1, tree depths 3, 5, and 10, crossed with trees specified 25 to 500 by 25 increments, random forest with trees specified 25 to 500 by 25 increments, k-nearest neighbor of specification 3, 4, 5, 7, and 9 number of nearest neighbors, k. Cases with *fitted* prognostic score include an augmented library that includes candidate learners with prognostic score in addition to without.

**Discrete super learner specifications** loss function is specified to be the mean log likelihood loss.

**Selected prognostic model** The selected learner from 20-fold cross validation is a linear regression model for both the reanalyses with $n = 419$ and $n = 100$.

# Appendix B: Expectation calculation when incorporating unobserved covariate

**Expectation calculation when incorporating unobserved covariate**

By definition, an unobserved covariate $U$ is never seen in real data but we include such a variable in simulation. We aim to demonstrate that even if the outcome model can be learned perfectly from historical data, if there exists an unobserved shift between the historical and trial sample, then the learned historical outcome model (prognostic model) can never be equivalent to the trial outcome model. We explicitly write out the expectation of the outcome Y given treatment A, baseline covariates W, and data set indicator D using an unobserved covariate U.

$$E[Y|\ A, W, D = d] = \int E[Y|\ A, W, U = u, D = d] p(u|A, W, D = d) \, \mathrm{d}u$$

A shift in the distribution $p(u|A, W, D = 1) \neq p(u|A, W, D = 0)$ of the unobserved covariate U will generally result in unequal conditional expectations, i.e., $E[Y|\ A, W, D = 1] \neq E[Y|\ A, W, D = 0]$.

This shift is the basis of the simulation for Figure 3 B. In our simulation, we have $U|A, W, D = 1 \sim \mathrm{Unif}(0, 1)$ and $U|A, W, D = 0 \sim \mathrm{Unif}(\underline{u}, \overline{u})$ where the limits $\underline{u}, \overline{u}$ increase or decrease past 0 or 1 depending on the desired magnitude of covariate shift. The "oracle" prognostic score is given by $E[Y|\ A, W, D = 1]$, i.e. always integrating over the correct (trial) density $U|A, W, D = 1 \sim \mathrm{Unif}(0, 1)$.

By framing shifts in the conditional mean as shifts in an unobserved covariate we can directly control the magnitude of the change instead of manually specifying different conditonal mean functions.

# Supplemental Appendix C: Simulation results for different data generation processes

| Scenario | Estimators *prog.* | Bias | Var. | SE bias | SE var. | RMSE | power | coverage |
|---|---|---|---|---|---|---|---|---|
| heterogeneous effect | TMLE *none* | 0.009 | 5.691 | 0.180 | 0.055 | 2.380 | 0.645 | 0.960 |
| | TMLE *fitted* | -0.064 | 4.482 | 0.158 | 0.005 | 2.113 | 0.720 | 0.975 |
| | TMLE *oracle* | -0.069 | 4.341 | 0.152 | 0.003 | 2.080 | 0.745 | 0.985 |
| | linear *none* | 0.009 | 9.774 | 0.213 | 0.026 | 3.119 | 0.405 | 0.950 |
| | linear *fitted* | -0.064 | 9.710 | 0.137 | 0.029 | 3.108 | 0.420 | 0.945 |
| | linear *oracle* | -0.069 | 9.727 | 0.138 | 0.028 | 3.111 | 0.420 | 0.945 |
| | unadjusted *none* | 0.153 | 9.509 | 0.139 | 0.011 | 3.080 | 0.435 | 0.950 |
| constant effect | TMLE *none* | 0.027 | 0.119 | -0.010 | 0.001 | 0.345 | 0.655 | 0.945 |
| | TMLE *fitted* | 0.025 | 0.067 | 0.028 | 0.000 | 0.260 | 0.790 | 0.970 |
| | TMLE *oracle* | 0.034 | 0.074 | 0.005 | 0.000 | 0.273 | 0.780 | 0.960 |
| | linear *none* | 0.032 | 0.427 | 0.014 | 0.001 | 0.653 | 0.240 | 0.940 |
| | linear *fitted* | 0.023 | 0.069 | 0.014 | 0.000 | 0.263 | 0.800 | 0.970 |
| | linear *oracle* | 0.025 | 0.060 | 0.020 | 0.000 | 0.246 | 0.840 | 0.970 |
| | unadjusted *none* | 0.067 | 0.857 | -0.037 | 0.001 | 0.926 | 0.150 | 0.945 |
| small observed shift | TMLE *none* | -0.001 | 5.699 | 0.178 | 0.055 | 2.381 | 0.640 | 0.960 |
| | TMLE *fitted* | -0.096 | 4.851 | 0.187 | 0.038 | 2.199 | 0.715 | 0.970 |
| | TMLE *oracle* | -0.074 | 4.411 | 0.133 | 0.003 | 2.096 | 0.740 | 0.985 |
| | linear *none* | 0.071 | 9.771 | 0.213 | 0.025 | 3.119 | 0.405 | 0.950 |
| | linear *fitted* | 0.094 | 9.835 | 0.205 | 0.026 | 3.130 | 0.405 | 0.950 |
| | linear *oracle* | 0.025 | 9.702 | 0.142 | 0.028 | 3.107 | 0.415 | 0.945 |
| | unadjusted *none* | 0.153 | 9.509 | 0.139 | 0.011 | 3.080 | 0.435 | 0.950 |
| small unobserved shift | TMLE *none* | -0.032 | 5.803 | 0.155 | 0.050 | 2.403 | 0.640 | 0.950 |
| | TMLE *fitted* | -0.026 | 5.346 | 0.108 | 0.015 | 2.306 | 0.695 | 0.975 |
| | TMLE *oracle* | -0.076 | 4.434 | 0.129 | 0.003 | 2.102 | 0.745 | 0.985 |
| | linear *none* | 0.058 | 9.992 | 0.174 | 0.025 | 3.154 | 0.390 | 0.940 |
| | linear *fitted* | -0.039 | 9.998 | 0.106 | 0.027 | 3.154 | 0.410 | 0.935 |
| | linear *oracle* | -0.022 | 9.774 | 0.128 | 0.027 | 3.119 | 0.390 | 0.935 |
| | unadjusted *none* | 0.118 | 9.314 | 0.173 | 0.010 | 3.046 | 0.425 | 0.950 |
| small historical sample $(\tilde{n}, n) = (100, 250)$ | TMLE *none* | 0.143 | 7.301 | -0.154 | 0.039 | 2.699 | 0.690 | 0.930 |
| | TMLE *fitted* | 0.121 | 6.006 | -0.074 | 0.017 | 2.448 | 0.735 | 0.925 |
| | TMLE *oracle* | 0.104 | 5.208 | -0.046 | 0.004 | 2.279 | 0.755 | 0.935 |
| | linear *none* | 0.234 | 11.884 | -0.097 | 0.027 | 3.447 | 0.440 | 0.935 |
| | linear *fitted* | 0.090 | 11.499 | -0.093 | 0.028 | 3.384 | 0.430 | 0.955 |
| | linear *oracle* | 0.093 | 11.309 | -0.091 | 0.027 | 3.356 | 0.425 | 0.950 |
| | unadjusted *none* | 0.146 | 10.439 | -0.009 | 0.010 | 3.226 | 0.425 | 0.960 |
| small trial sample $(\tilde{n},\ n) =$ (1000, 100) | TMLE *none* | 0.758 | 24.859 | 0.366 | 0.237 | 5.031 | 0.215 | 0.960 |
| | TMLE *fitted* | 0.155 | 14.698 | -0.116 | 0.039 | 3.827 | 0.380 | 0.955 |
| | TMLE *oracle* | 0.092 | 14.016 | -0.083 | 0.041 | 3.736 | 0.365 | 0.960 |
| | linear *none* | 0.647 | 30.222 | 0.208 | 0.285 | 5.522 | 0.200 | 0.940 |
| | linear *fitted* | 0.607 | 30.267 | 0.080 | 0.336 | 5.521 | 0.210 | 0.945 |
| | linear *oracle* | 0.620 | 30.279 | 0.088 | 0.326 | 5.524 | 0.205 | 0.950 |
| | unadjusted *none* | 0.670 | 21.590 | 0.468 | 0.076 | 4.683 | 0.205 | 0.970 |

Supplemental Appendix Table 5: Mean of empirically estimated bias, variance, and standard errors of them for the targeted maximum likelihood estimator with or without prognostic score across different DGPs. For all the scenarios the conditional means are shared with the heterogeneous effect DGP, except for the constant effect DGP. Unless otherwise specified $(\tilde{n}, n) = (1000, 250)$.

# Appendix D: Historical sample summary

| Trial ID | Duration | Titration target | Number of subjects on insulin IGlar | |
| --- | --- | --- | --- | --- |
| | | | Randomised | Completed |
| NN9068-4228 | 104 weeks | 4.0-5.0 mmol/L | 506 | 481 |
| NN1250-3579 | 52 weeks | 4.0-5.0 mmol/L | 257 | 197 |
| NN1250-3586 | 26 weeks | 4.0-5.0 mmol/L | 146 | 136 |
| NN1250-3587 | 26 weeks | 4.0-5.0 mmol/L | 278 | 254 |
| NN1250-3672 | 26 weeks | 4.0-5.0 mmol/L | 230 | 201 |
| NN1250-3718 | 26 weeks | 4.0-5.0 mmol/L | 234 | 209 |
| NN1250-3724 | 26 weeks | 4.0-5.0 mmol/L | 230 | 206 |
| NN9535-3625 | 30 weeks | 4.0-5.5 mmol/L | 365 | 343 |
| NN2211-1697 | 26 weeks | $\leq 5.0$ mmol/L | 234 | 219 |
| NN5401-3590 | 26 weeks | 3.9-5.0 mmol/L | 264 | 232 |
| NN5401-3726 | 26 weeks | 3.9-5.0 mmol/L | extension of 3590 | 209 |
| NN5401-3896 | 26 weeks | 3.9-5.0 mmol/L | 149 | 137 |
| NN1436-4383 | 26 weeks | 4.4-7.2 mmol/L | 122 | 119 |
| NN1436-4465 | 16 weeks | 4.4-7.2 mmol/L | 51 | 51 |
| NN1436-4477 | 78 weeks | 4.4-7.2 mmol/L | 492 | 477 |

Supplemental Appendix Table 6: Summary of the trials that constitute the historical sample.

# Appendix E: Summary of continuous measurements of the baseline covariates in the case study

|  | Historical sample | New RCT sample |
|---|---|---|
| Number of subjects | 3311 | 419 |
| Age (years) | | |
|   N | 3311 | 419 |
|   Mean (SD) | 57.34 (9.92) | 56.67 (10.28) |
|   Median | 58.00 | 58.00 |
|   Min ; Max | 21.00; 85.00 | 25.00; 83.00 |
| Alanine aminotransferase (U/L) | | |
|   N | 3303 | 419 |
|   Mean (SD) | 29.51 (17.97) | 26.63 (15.48) |
|   Median | 25.00 | 23.00 |
|   Min ; Max | 2.50; 333.00 | 6.00; 138.00 |
| Albumin (g/dL) | | |
|   N | 3306 | 419 |
|   Mean (SD) | 4.48 (0.28) | 4.51 (0.25) |
|   Median | 4.50 | 4.50 |
|   Min ; Max | 2.50; 5.90 | 3.80; 5.20 |
| Alkaline phosphatase (U/L) | | |
|   N | 3305 | 419 |
|   Mean (SD) | 75.99 (23.50) | 71.82 (22.31) |
|   Median | 73.00 | 68.00 |
|   Min ; Max | 19.00; 261.00 | 20.00; 196.00 |
| Aspartate aminotransferase (U/L) | | |
|   N | 3299 | 419 |
|   Mean (SD) | 23.22 (12.13) | 21.38 (9.86) |
|   Median | 20.00 | 19.00 |
|   Min ; Max | 6.00; 227.00 | 6.00; 89.00 |
| Basophils Blood (%) | | |
| N | 3284 | 419 |
| Mean (SD) | 0.53 (0.38) | 0.39 (0.22) |
| Median | 0.40 | 0.40 |
| Min ; Max | 0.00; 4.40 | 0.00; 1.60 |
| BMI (kg/m^2) | | |
| N | 3309 | 419 |
| Mean (SD) | 30.72 (5.69) | 31.22 (4.82) |
| Median | 30.22 | 31.00 |
| Min ; Max | 16.01; 56.39 | 20.00; 43.30 |
| Body weight (kg) | | |
| N | 3309 | 419 |
| Mean (SD) | 86.27 (19.82) | 88.30 (17.41) |
| Median | 84.90 | 86.30 |
| Min ; Max | 36.30; 171.70 | 50.98; 145.33 |
| Change from baseline to week 26 HbA1c | | |
| N | 2642 | 399 |
| Mean (SD) | -1.48 (1.01) | -1.81 (1.01) |
| Median | -1.40 | -1.70 |
| Min ; Max | -5.30; 2.60 | -6.20 ; 1.20 |
| Creatinine (umol/L) | | |
| N | 3308 | 419 |
| Mean (SD) | 74.14 (18.56) | 73.60 (15.64) |
| Median | 72.00 | 72.00 |
| Min ; Max | 23.00; 409.00 | 36.00; 121.00 |
| Diabetes duration (years) | | |
| N | 3311 | 419 |
| Mean (SD) | 9.67 (6.36) | 9.55 (6.26) |
| Median | 8.67 | 8.47 |
| Min ; Max | 0.30; 49.65 | 0.44; 34.24 |
| Diastolic Blood Pressure (mmHg) | | |
| N | 3309 | 419 |
| Mean (SD) | 78.85 (8.53) | 79.12 (8.37) |
| Median | 80.00 | 80.00 |
| Min ; Max | 47.00; 116.00 | 57.00; 109.00 |
| Eosinophils Blood (%) | | |
| N | 3284 | 419 |
| Mean (SD) | 2.65 (2.35) | 2.56 (2.06) |
| Median | 2.10 | 2.00 |
| Min ; Max | 0.00; 43.70 | 0.00; 15.20 |
| Erythrocytes (10^12/L) | | |
| N | 3297 | 419 |
| Mean (SD) | 4.68 (0.45) | 5.08 (0.48) |
| Median | 4.70 | 5.00 |
| Min ; Max | 3.10; 7.40 | 3.60; 7.50 |
| FPG (mmol/L) | | |
| N | 3268 | 411 |
| Mean (SD) | 9.81 (2.65) | 9.55 (2.53) |
| Median | 9.50 | 9.20 |
| Min ; Max | 2.70; 22.60 | 3.60; 29.20 |
| Haematocrit Blood (%) | | |
| N | 3264 | 419 |
| Mean (SD) | 42.49 (4.16) | 45.28 (4.24) |
| Median | 42.50 | 45.50 |
| Min ; Max | 22.80; 60.50 | 31.20; 58.40 |
| Haemoglobin Blood (mmol/L) | | |
| N | 3297 | 419 |
| Mean (SD) | 8.63 (0.88) | 8.99 (0.88) |
| Median | 8.68 | 9.05 |
| Min ; Max | 4.34; 12.09 | 5.95; 11.16 |
| HbA1c at baseline (%) | | |
| N | 3311 | 419 |
| Mean (SD) | 8.39 (0.92) | 8.28 (1.01) |
| Median | 8.30 | 8.10 |
| Min ; Max | 6.60; 12.80 | 6.50; 13.50 |
| HDL cholesterol (mmol/L) | | |
| N | 3277 | 410 |
| Mean (SD) | 1.18 (0.33) | 1.21 (0.35) |
| Median | 1.14 | 1.14 |
| Min ; Max | 0.21; 3.99 | 0.31; 2.69 |
| Height (m) | | |
| N | 3311 | 419 |
| Mean (SD) | 1.67 (0.10) | 1.68 (0.09) |
| Median | 1.67 | 1.68 |
| Min ; Max | 1.36; 2.03 | 1.43; 2.01 |
| LDL cholesterol (mmol/L) | | |
| N | 3270 | 409 |
| Mean (SD) | 2.46 (0.94) | 2.42 (1.01) |
| Median | 2.36 | 2.28 |
| Min ; Max | 0.00; 6.73 | 0.10; 7.10 |
| Leukocytes (10^9/L) | | |
| N | 3297 | 419 |
| Mean (SD) | 7.33 (1.93) | 7.93 (2.04) |
| Median | 7.10 | 7.80 |
| Min ; Max | 2.80; 17.60 | 3.60; 15.80 |
| Lymphocytes Blood (%) | | |
| N | 3284 | 419 |
| Mean (SD) | 30.00 (7.88) | 29.39 (7.66) |
| Median | 29.70 | 28.70 |
| Min ; Max | 4.60; 71.00 | 10.70; 55.10 |
| Monocytes Blood (%) | | |
| N | 3284 | 419 |
| Mean (SD) | 5.88 (2.19) | 5.83 (2.30) |
| Median | 5.70 | 5.70 |
| Min ; Max | 0.00; 21.70 | 0.50; 17.20 |
| Neutrophils Blood (%) | | |
| N | 3284 | 419 |
| Mean (SD) | 60.93 (8.68) | 61.83 (9.00) |
| Median | 61.10 | 62.20 |
| Min ; Max | 16.60; 91.60 | 25.20; 86.50 |
| Potassium (mmol/L) | | |
| N | 3304 | 419 |
| Mean (SD) | 4.48 (0.42) | 4.53 (0.41) |
| Median | 4.49 | 4.50 |
| Min ; Max | 3.10; 7.00 | 3.30; 6.50 |
| Pulse (beats/min) | | |
| N | 3310 | 419 |
| Mean (SD) | 75.31 (10.00) | 75.56 (9.34) |
| Median | 75.00 | 76.00 |
| Min ; Max | 45.50; 118.00 | 52.00; 108.00 |
| Sodium (mmol/L) | | |
| N | 3303 | 419 |
| Mean (SD) | 139.73 (2.81) | 140.19 (2.44) |
| Median | 140.00 | 140.00 |
| Min ; Max | 121.00; 154.00 | 132.00; 148.00 |
| Systolic Blood Pressure (mmHg) | | |
| N | 3309 | 419 |
| Mean (SD) | 131.53 (14.40) | 129.69 (13.69) |
| Median | 131.00 | 130.00 |
| Min ; Max | 90.00; 200.00 | 96.00; 171.00 |
| Thrombocytes (10^9/L) | | |
| N | 3269 | 419 |
| Mean (SD) | 240.18 (64.34) | 244.27 (64.91) |
| Median | 233.00 | 242.00 |
| Min ; Max | 13.00; 611.00 | 63.00; 477.00 |
| Total bilirubin (umol/L) | | |
| N | 3304 | 419 |
| Mean (SD) | 8.10 (4.33) | 8.12 (4.73) |
| Median | 7.00 | 7.00 |
| Min ; Max | 0.00; 36.00 | 1.00; 33.00 |
| Total cholesterol (mmol/L) | | |
| N | 3284 | 410 |
| Mean (SD) | 4.54 (1.13) | 4.59 (1.28) |
| Median | 4.43 | 4.43 |
| Min ; Max | 0.93; 13.93 | 2.02; 11.37 |
| Triglycerides (mmol/L) | | |
| N | 3279 | 410 |
| Mean (SD) | 2.07 (1.78) | 2.23 (2.36) |
| Median | 1.65 | 1.70 |
| Min ; Max | 0.24; 34.25 | 0.38; 27.80 |

Supplemental Appendix Table 7: Summary of the continous baseline covariates. N: number of participants with non missing values. SD: standard deviation. BMI: body mass index. FPG: fasting plasma glucose. HbA1C: hemoglobin A1C. HDL: high density lipoprotein. LDL: low density lipoprotein.

# Appendix F: Summary of categorical measurements of the baseline covariates in the case study

|  | Historical sample | | New RCT sample | |
|---|---|---|---|---|
|  | N | (%) | N | (%) |
| Number of subjects | 3311 |  | 419 |  |
| Sex |  |  |  |  |
|   Female | 1480 | (44.7) | 173 | (41.3) |
|   Male | 1831 | (55.3) | 246 | (58.7) |
| Race |  |  |  |  |
|   Asian | 820 | (24.8) | 65 | (15.5) |
|   Black or African American | 183 | (5.5) | $< 5$ |  |
|   Other | 67 | (2.0) | $< 5$ |  |
|   White | 2241 | (67.7) | 346 | (82.6) |
| Smoking status |  |  |  |  |
|   Current | 241 | (7.3) | 53 | (12.6) |
|   Never | 910 | (27.5) | 249 | (59.4) |
|   Previous | 378 | (11.4) | 116 | (27.7) |
| Region |  |  |  |  |
|   Asia | 748 | (22.6) | 57 | (13.6) |
|   Europe | 1303 | (39.4) | 228 | (54.4) |
|   North america | 1015 | (30.7) | 89 | (21.2) |
|   South africa | 91 | (2.7) | 0 |  |
|   South america | 154 | (4.7) | 45 | (10.7) |
| Ethnicity |  |  |  |  |
|   Hispanic or Latino | 461 | (13.9) | 68 | (16.2) |
|   Not Hispanic or Latino | 2806 | (84.7) | 351 | (83.8) |
| Titration target |  |  |  |  |
|   3.9-5.0 mmol/l | 410 | (12.4) | 0 |  |
|   4.0-5.0 mmol/l | 2236 | (67.5) | 419 | (100.0) |
|   4.4-7.2 mmol/l | 665 | (20.1) | 0 |  |
| Blinding | | | | |
|   Double-blinded | 122 | (3.7) | 0 | (16.2) |
|   Open-label | 3189 | (96.3) | 419 | (100.0) |
| Biguanides | | | | |
|   Yes (continued in trial) | 2873 | (86.8) | 369 | (88.1) |
|   Yes (discontinued in trial) | 248 | (7.5) | 27 | (6.4) |
|   No | 190 | (5.7) | 23 | (5.5) |
| Sulfonylureas | | | | |
|   Yes (continued in trial) | 436 | (13.2) | < 5 | |
|   Yes (discontinued in trial) | 1485 | (44.9) | 0 | |
|   No | 1390 | (42.0) | >414 | |
| DPP4 | | | | |
|   Yes (continued in trial) | 278 | (8.4) | < 5 | |
|   Yes (discontinued in trial) | 361 | (10.9) | > 123 | |
|   No | 2672 | (80.7) | > 285 | |
| Other blood glucose lowering drugs | | | | |
|   Yes (continued in trial) | < 5 | | 0 | |
|   Yes (discontinued in trial) | > 79 | | 0 | |
|   No | > 3220 | | 419 | (100.0) |
| Alpha Glucosidase inhibitor | | | | |
|   Yes (continued in trial) | 87 | (2.6) | 0 | |
|   Yes (discontinued in trial) | 69 | (2.1) | 0 | |
|   No | 3155 | (95.3) | 419 | (100.0) |
| Combination of blood glucose lowering drug | | | | |
|   Yes (continued in trial) | 32 | (1.0) | 0 | |
|   Yes (discontinued in trial) | 36 | (1.1) | 8 | (1.9) |
|   No | 3243 | (97.9) | 411 | (98.1) |

| | | | | | |
|---|---|---|---|---|---|
| Thiazolidinediones | | | | | |
|   Yes (continued in trial) | 78 | (2.4) | | 20 | (4.8) |
|   Yes (discontinued in trial) | 29 | (0.9) | | 0 | |
|   No | 3204 | (96.8) | | 399 | (95.2) |
| SGLT2i | | | | | |
|   Yes (continued in trial) | 175 | (5.3) | | > 383 | |
|   Yes (discontinued in trial) | 15 | (0.5) | | > 25 | |
|   No | 3121 | (94.3) | | < 5 | |
| GLP-1 receptor agonist | | | | | |
|   Yes (continued in trial) | 79 | (2.4) | | 0 | |
|   Yes (discontinued in trial) | 13 | (0.4) | | 0 | |
|   No | 3219 | (97.2) | | 419 | (100.0) |

Supplemental Appendix Table 8: Categorical baseline covariates of the case study

N: number of participants. The symbol < and > are used to anonymise the number of participants in the category, for example, when the number is lower than 5.

# Appendix G: Missing pattern of the case study

To clean and currate the 14 data sets we imputed the HbA1C at week 26 value. For the historical sample the imputation was made using an ANCOVA model with last observed HbA1C measurement before landmark visit, time point of last measurement, baseline HbA1C, discontinuation prior to week 26 indicator and study-id as adjustment covariates. For the new trial data a similar approach was employed. However, in this case the last observed HbA1C measurement before landmark visit week 26, time point of last measurement, baseline HbA1C, discontinuation prior to week 26 indicator, region, treatment indicator and pre-study OADs were used as adjustment covariates. This was done in order to use a similar imputation as used in the original analysis.

After imputing the HbA1C at week 26 value a total 94.4% of the participants had complete data for the combined historical and new trial data. The missingness of the covariates is displayed below. For the covariates we included missingness indicators and respectively imputed covariates using random forest [44]. This was done seperately on the historical and new trial data. The normalized root mean square error was 0.218 for continuous covariates and proportion of falsely classified is 0.004 for the historical data sample. The normalized root mean square error was 0.010 for continuous covariates and proportion of falsely classified is 0.023 for the new trial data. The missingness indicators of the historical sample did all overlap with the missingness indicators from NN9068-4229 trial. Since some of the covariates had near zero variance, were colinear or had large absolute correlation with each other we removed some of the covariates.

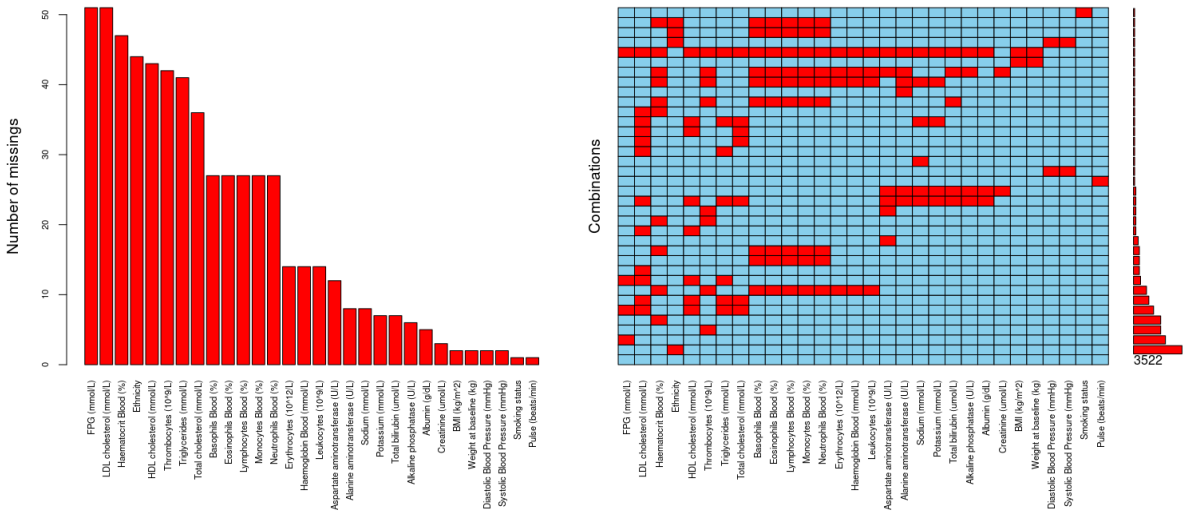

Figure 4: Left: Number of missing covariates. Right: Combination of missingness of covariates.

Due to near 0 variance, coliniarity or high absolute correlation between the covariates, we excluded some of the values. Thus the baseline covariates used in the model where reduced to the following:

- Age
- Diabetes duration
- BMI
- HbA1C
- Height
- Weight
- Alanine aminotransferase
- Albumin
- Alkaline phosphatase
- Aspartate aminotransferase
- Basophils
- Creatinine
- Eosinophils
- Erythrocytes
- Fasting plasma glucose
- Haematocrit
- HDL cholesterol
- LDL cholesterol
- Leukocytes
- Lymphocytes
- Monocytes
- Potassium
- Sodium
- Thrombosytes
- Total bilirubin
- Total cholesterol
- Triglycerides
- Diastolic blood pressure
- Pulse
- Systolic blood pressure
- Country
- Sex
- Race
- Smoking status
- Region
- Ethnicity
- Biguanides
- DPP4
- SGLT2I
- Previos OADs